\documentclass[aps,showpacs,superscriptaddress,preprint]{revtex4}
\usepackage{amsfonts}
\usepackage{pifont}
\usepackage{mathrsfs}
\usepackage{amssymb}
\usepackage{graphicx,epsfig,latexsym,overpic,amssymb,color}
\date{\today}
\begin{document}
\title{Shape coexistence and triaxiality in nuclei near $^{80}$Zr}
\author{S.J. Zheng}
\affiliation{School of Physics and State Key Laboratory of Nuclear
Physics and Technology, Peking University, Beijing 100871, China}
\author{F.R. Xu}
\email{frxu@pku.edu.cn} \affiliation{School of Physics and State Key
Laboratory of Nuclear Physics and Technology, Peking University,
Beijing 100871, China}\affiliation{Center of Theoretical Nuclear
Physics, National Laboratory of Heavy Ion Accelerator of Lanzhou,
Lanzhou 730000, China}
\author{S.F. Shen}
\affiliation{School of Physics and State Key Laboratory of Nuclear
Physics and Technology, Peking University, Beijing 100871, China}
\affiliation{Institute of Nuclear Energy Safety Technology, Chinese
Academy of Sciences, Hefei 230031, Anhui, China}\affiliation{School
of Physics, Suranaree University of Technology, Nakhon Ratchasima
30000, Thailand}
\author{H.L. Liu}
\affiliation{School of Physics and State Key Laboratory of Nuclear
Physics and Technology, Peking University, Beijing 100871, China}
\author{R. Wyss}
\affiliation{AlbaNova University Center, Royal Institute of
Technology, S-106 91 Stockholm, Sweden}

\begin{abstract}
Total-Routhian-Surface calculations have been performed to
investigate the shape evolutions of $A\sim80$ nuclei, $^{80-84}$Zr,
$^{76-80}$Sr and $^{84,86}$Mo. Shape coexistences of spherical,
prolate and oblate deformations have been found in these nuclei.
Particularly for the nuclei, $^{80}$Sr and $^{82}$Zr, the energy
differences between two shape-coexisting states are less than 220
keV. At high spins, the $g_{9/2}$ shell plays an important role for
shape evolutions. It has been found that the alignment of the
$g_{9/2}$ quasi-particles drives nuclei to be triaxial.
\end{abstract}
\pacs{ 21.10.-k, 21.60.-n, 27.50.+e} \maketitle
\section{introduction}

The $A\sim80$ nuclei far from the $\beta$-stability line are
attracting significant attention because of various shape evolutions
and shape coexistences. Moreover, they locate at the key points in
the $rp$ (rapid proton capture) process path, playing an important
role in the astrophysical nuclear synthesization. According to
mean-field models, single-particle level densities in this mass
region are noticeably low. There are significant shell gaps at the
prolate deformation $\beta_2\approx0.4$ with the particle number 38
or 40 and at oblate deformation $\beta_2\approx-0.3$ with the
particle number 34 or 36, which leads to rich shape transitions with
changing nucleon numbers. Up to date, the shape coexistence of
prolate, oblate and triaxial deformations has been seen in
$^{82}$Sr~\cite{Baktash}. Superdeformed bands in $^{80-83}$Sr,
$^{82-84}$Y and $^{83,84,86}$Zr have been established experimentally
(see Ref.~\cite{Lerma} and references therein). Some of
superdeformed bands in $^{86}$Zr~\cite{Sara} and
$^{80}$Sr~\cite{Devlin} are suggested to be triaxial. Theoretical
calculations predicted that $^{80}$Sr and $^{84}$Zr have triaxial
deformations~\cite{Naz,Dudek}. Also global calculations with the
axial asymmetry shape have been done recently for the ground states
of the nuclei~\cite{Moller06}. The $A\sim80$ nuclei lie in the
region where axial symmetry is broken usually. However, it is not
easy experimentally to deduce the information about the triaxial
deformation. The wobbling, which has been observed in
$^{163,165,167}$Lu~\cite{Jen,Sch,Amro}, has been considered to be
the proof of triaxiality. With the advance of experimental
techniques, detailed observations in this mass region become
available.

It is predicted that $A\sim80$ nuclei could contain higher-order
geometrical symmetry, such as possible tetrahedral deformation in
ground and low excited states~\cite{Dudek02}. The exotic deformation
correlation comes from the special shell structure~\cite{Dudek02},
giving rich shape information. Though there could be many elements
(e.g., pairing correlation) affecting the deformation, the shell
structure in the deformed potential governs the shape development of
nuclear states. In the present work, we focus on the deformation
evolutions and shape coexistences in the $A\sim80$ nuclei.

\section{the model}

The Total-Routhian-Surface (TRS) calculations~\cite{Naz89} have been
performed. The total Routhian $E^\omega(Z,N,\hat\beta)$ of a nucleus
$(Z,N)$ at a rotational frequency $\omega$ and deformation
$\hat\beta$ is calculated as follows~\cite{Naz89}
\begin{equation}
\begin{array}{ll}
E^\omega(Z,N,\hat\beta)= & E^{\omega=0}(Z,N,\hat\beta)+
[<\Psi^\omega|\hat{H}^\omega|\Psi^\omega> \\
& -<\Psi^\omega|\hat{H}^\omega|\Psi^\omega>_{\omega=0}],\\
\end{array}
\end{equation}
where $E^{\omega=0}(Z,N,\hat\beta)$ is the total energy at zero
frequency, consisting of the macroscopic liquid-drop
energy~\cite{Myers}, the microscopic shell
correction~\cite{Naz90,Strutinsky} and pairing energy~\cite{Satula}.
The last two terms in the bracket represent the change in energy due
to the rotation. The total Hamiltonian is written as~\cite{Naz89}
\begin{equation}
\begin{array}{ll}
\hat{H}^\omega = & \sum\limits_{ij}[(<i|h_{ws}|j>-
\lambda\delta_{ij})a_i^+a_j
 - \omega<i|\hat{j}_x|j>a_i^+a_j] \\
& - G\sum\limits_{i,i'>0}a_i^+a_{\bar{i}}^+a_{i'}a_{\bar{i'}}.\\
\end{array}
\end{equation}
For the single-particle Hamiltonian, $h_{ws}$, a non-axial deformed
Woods-Saxon (WS) potential has been adopted.

The pairing is treated by the Lipkin-Nogami approach~\cite{Satula}
in which the particle number is conserved approximately and thus the
spurious pairing phase transition encountered in the BCS calculation
can be avoided (see Ref.~\cite{Satula} for the detailed formulation
of the cranked Lipkin-Nogami TRS method). Both monopole and
quadruple pairings are considered~\cite{Satula prc} with the
monopole pairing strength $G$ determined by the average gap
method~\cite{Moller92} and quadruple strengths obtained by restoring
the Galilean invariance broken by the seniority pairing
force~\cite{Saka,Satula prc,Xu}. The TRS calculation is performed in
the deformation space $\hat\beta = (\beta_2, \gamma, \beta_4)$.
Pairing correlations are dependent on the rotational frequency and
deformation. In order to include such dependence in the TRS, we have
run the pairing-deformation-frequency self-consistent TRS
calculation, i.e., for any given deformation and frequency, pairing
is self-consistently treated by the Hartree-Fock-Bogolyubov-like
equation~\cite{Satula}. At a given frequency, the deformation of a
state is determined by minimizing the calculated TRS.

For $N = Z$ nuclei, the neutron-proton pairing can be
remarkable~\cite{Satula plb,Wyss}. In the present work, however, we
are interested in the shape of the nuclei. The pairing correlation
should not affect the deformation significantly. The shell
correction dominates the shape of a state.


\section{CALCULATIONS AND DISCUSSIONS}

The TRS calculations for even-even Sr, Zr, Mo isotopes near $A=80$
have been performed. Deformations, shape coexistences and collective
rotational properties with increasing rotational frequency are
discussed. The kinematic moments of inertia are calculated and
compared with existing experimental data.

\subsection{Deformations and shape coexistences in ground and low-lying states}

\begin{figure}
\centering
\includegraphics[width=0.9\textwidth]{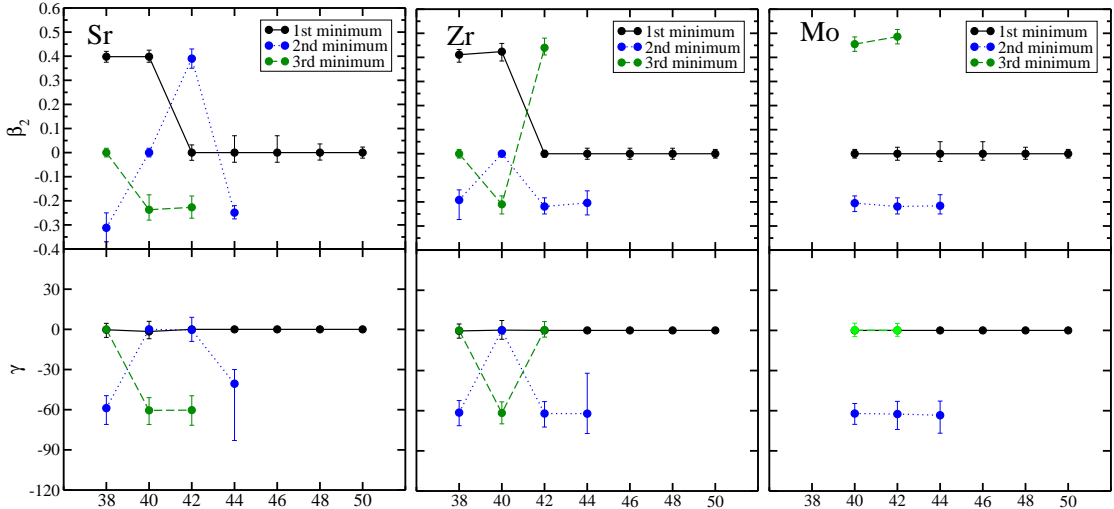}
\caption{Calculated quadrupole deformations $\beta_2$ and $\gamma$
for even-even Sr, Zr and Mo isotopes with $38\leqslant
N\leqslant50$. The error bar displays the deformation values within
an energy range of less than 100 keV above the minimum, giving an
indication of the softness of the nucleus with respect to the
corresponding shape parameter. The black dots represent ground state
deformations (1st minimum). The blue and green dots represent the
coexisting deformations (2nd and 3rd minimum).} \label{fig:1}
\end{figure}

Due to the large shell gaps at prolate, oblate deformations, there
is a competition between different deformations. The calculated
deformations with ($\beta$,$\gamma$) of ground states and excited
states for even-even Sr, Zr, Mo isotopes are shown in
Fig.~\ref{fig:1}. Softnesses which indicate the stability of
deformations are also calculated and shown. Various shape coexisting
states with spherical, prolate, oblate and triaxial deformations are
found in this mass region.

TRS calculations show that the nuclei with $N\leq 40$ have
well-deformed ($\beta_2\approx$ 0.4) prolate ground states, which is
consistent with the experimental data, $\epsilon_2=0.4$ and
$\epsilon_2=0.39$
($\epsilon_2\approx0.944\beta_2-0.122\beta_{2}^{2}+0.154\beta_{2}\beta_{4}-0.199\beta_{4}^{2}$)
for the nuclei $^{78}$Sr~\cite{Lister-sr} and
$^{80}$Zr~\cite{Lister-zr}, respectively. With the neutron number
$N$ approaching to the magic number 50, the shapes of the nuclei
become spherical. The transitional nuclei between the well-deformed
and spherical nuclei are quite soft. Experimentally, the $\gamma$
vibrational bands have been observed for $^{80,82}$Sr, $^{82,84}$Zr
and $^{86}$Mo~\cite{Sienko,Tabor,Rudolph,Doring,Andgren} indicating
the soft deformations. The energy ratios of E($4^+$)/E($2^+$), which
indicate the degrees of the collectivity, show the similar trend.
They decrease steadily from more than 2.80, through 2.5 to 2.3 for
N=40, 42, 44 isotones respectively, showing the decreasing
collectivity.

%

For strontium isotopes, a shape transition of the ground-state
deformations occurs from prolate at $N=40$ to spherical shapes at
$N=42$. Due to the large shell gaps at prolate, oblate and spherical
shapes, there is a competition between different deformations.
Consequently, prolate, oblate and spherical shapes coexist as shown
in the top panel of Fig.~\ref{fig:1}. The nuclei $^{76,78}$Sr have
well-deformed prolate ground states with $\beta_2\approx 0.4$.
Shallow oblate deformations with $\beta_2=-0.31$ ($^{76}$Sr) and
$\beta_2=-0.24$ ($^{78}$Sr) coexist, which is by about 2.04 MeV and
1.53 MeV higher for $^{76}$Sr and $^{78}$Sr, respectively. The
nucleus $^{80}$Sr has a spherical ground state. However, it has very
low-lying states of prolate ($\beta_2=0.39$) and oblate
($\beta_2=-0.23$) deformations, which is only by 70 keV and 290 keV
higher than the ground state. A triaxial shape with
($\beta_2,\gamma$)=($0.25, -40^\circ$) exists in the nucleus
$^{82}$Sr, while it is rather soft in $\gamma$ direction from
$-80^\circ$ to $-30^\circ$. $\gamma$ vibrational band in $^{82}$Sr
has been observed~\cite{Tabor} indicating the soft $\gamma$
deformation.

A shape transition from prolate to spherical shapes also occurs for
the zirconium isotopes as shown in the middle panel of
fig.~\ref{fig:1}. Similar with the strontium isotopes, prolate,
oblate and spherical deformations coexist for the nuclei
$^{78,80,82}$Zr, while the energy differences between the prolate
and oblate deformations vary from 2.02 MeV through 1.28 MeV to
-0.112 MeV for $^{78}$Zr, $^{80}$Zr, $^{82}$Zr respectively.

For the molybdenum isotopes, spherical shapes become the ground
states, while coexistence of prolate or oblate deformation also
appears for the nuclei $^{82,84,86}$Mo as shown in the right panel
of fig.~\ref{fig:1}. For the $Z=N$ nucleus, $^{84}$Mo, little is
known about the shape information up to date. Several calculations
have predicted different shapes. M\"{o}ller and Nix~\cite{Moller81}
predicted an oblate deformation with $\varepsilon_2=-0.18$
($\varepsilon_2=0.95\beta_2$). Nazarewicz {\em et al}.~\cite{Naz}
found it to be soft axially symmetric. A nearly spherical ground
state is calculated by Aberg~\cite{Aberg}. Petrovici {\em et
al}.~\cite{petrovici} predicted it to be prolate with an oblate
deformation coexisting by above 3 MeV higher. In our calculations,
the ground state is spherical and an oblate shape coexists which is
only by 420 keV higher. Also a highly deformed prolate shape appears
with an even higher energy (more than 1.5 MeV).

\begin{figure}
\centering
\includegraphics[width=0.40\textwidth]{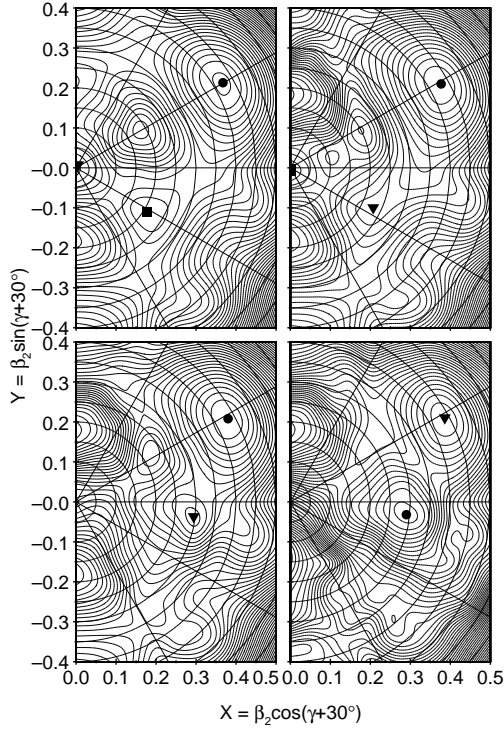}
\caption{The total Routhian surfaces for $^{80}$Zr at
$\hbar\omega$=0.0 MeV (upper-left), 0.45 MeV (upper-right), 0.55
(lower-left) and 0.7 MeV (lower-right). The filled dots represent
the first minima. The filled down-triangles and squares represent
the second and third minima, respectively. Contours are at a 200 keV
interval.} \label{fig:2}\end{figure}

\begin{figure}
\centering
\includegraphics[width=0.35\textwidth]{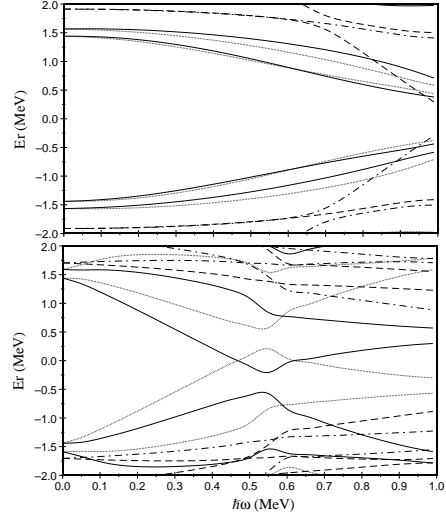}
\caption{Quasineutron Routhians of $^{80}$Zr for prolate (upper
panel) and triaxial (lower panel) deformations with $(\beta_2,
\gamma)=(0.424, 0.2^\circ)$ and $(0.297, -37.9^\circ)$,
respectively. ($\pi$, $\alpha$): solid=(+, +1/2), dotted=(+, -1/2),
dot-dash=(-, +1/2), dashed=(-, -1/2)} \label{fig:3}
\end{figure}

\subsection{triaxial deformations in neutron deficient Sr, Zr and Mo isotopes at high spins}


Experimental data for $^{80}$Zr is very sparse since it is
approaching the proton dripline, while its characters are crucial
which play an important role in the $rp$ process. Recently, five
cascade transitions have been verified with the ground-state band
established up to $10^+$ at 3789 keV~\cite{Fischer}. Alignment delay
has been predicted to appear at higher spins. Neutron-proton pairing
interaction may be the cause of the delayed rotational alignment.
However, the spin alignment is also sensitively influenced by other
factors such as deformation~\cite{Marginean}. Thus the shape
transitions are discussed in details for the nuclei of this region.


TRS diagrams for $^{80}$Zr are shown in Fig.~\ref{fig:2} at specific
rotational frequencies, $\hbar\omega$ = 0.0, 0.45, 0.55, and 0.70
MeV respectively. At $\hbar\omega=0$ MeV, that is, for the ground
state of the nucleus, a well-deformed prolate shape with
$\beta_2=0.42$ coexists with an oblate shape as discussed above. As
rotational frequency increases, the oblate deformation minimum
becomes shallower and it disappears at $\hbar\omega=0.45$ MeV. Then
a stable triaxial shape develops at $\hbar\omega$ = 0.55 MeV and its
total Routhian energy gets lower at higher rotational frequency. At
$\hbar\omega\approx$ 0.7 MeV it becomes yrast with $I\sim 19\hbar$.
Subsequently, the nucleus shows much more stable triaxiality up to
38$\hbar$. 

Subsequently, the single-quasiparticle Routhians are calculated and
shown in Fig.~\ref{fig:3} for the prolate and triaxial deformations.
For the $N=Z$ nucleus $^{80}$Zr, the neutrons and protons occupy the
same single particle orbitals and they have nearly the same
Routhians, therefore only quasi-neutron Routhians are shown. For the
prolate deformation, there is no alignments shown. For the triaxial
deformation, neutron alignments are predicted to emerge at the
rotational frequency about 0.55 MeV and proton alignments emerge
simultaneously. It is considered to be alignments of $g_{9/2}$
quasiprotons and quasineutrons as its neighboring nuclei $^{81}$Sr
and $^{84}$Zr~\cite{Moore,Dudek}. Note that triaxial deformation
develops after the alignments as shown in Fig. 2. That is to say,
the alignments of four $g_{9/2}$ quasi-particles drive the
nucleus to be triaxial. 
It is consistent with the prediction~\cite{Luo} that the alignment
of $g_{9/2}$ proton orbits tends to a positive $\gamma$-drive when
$Z$=40. Experimentally, it has been found that the $g_{9/2}$ neutron
polarizes the soft core of the nuclei
$^{55,57,59}$Cr~\cite{Deacon05} and $^{59}$Fe~\cite{Deacon07}. For
the nuclei in the A = 80 region, particularly for the nucleus
$^{80}$Zr, both $g_{9/2}$ proton and neutron orbits are occupied,
thus the shape driving effects being strengthened.

Calculations for the neighboring nuclei $^{76-80}$Sr, $^{82,84}$Zr
and $^{84,86}$Mo are also carried out with cranking TRS model,
showing a similar trend. The triaxially deformed shape develops
after the alignments of $g_{9/2}$ quasiprotons and quasineutrons.

The moments of inertia are very sensitive to the nuclear shape. Thus
the kinematic moments of inertia are calculated and compared with
experimental results as shown in Fig.~4-6, for $^{80-84}$Zr,
$^{76-80}$Sr and $^{84,86}$Mo respectively. The method of converting
the sequence of $\gamma$ rays into moments of inertia is referred to
Ref.~\cite{Ben}.

\begin{figure}
\includegraphics[width=0.35\textwidth]{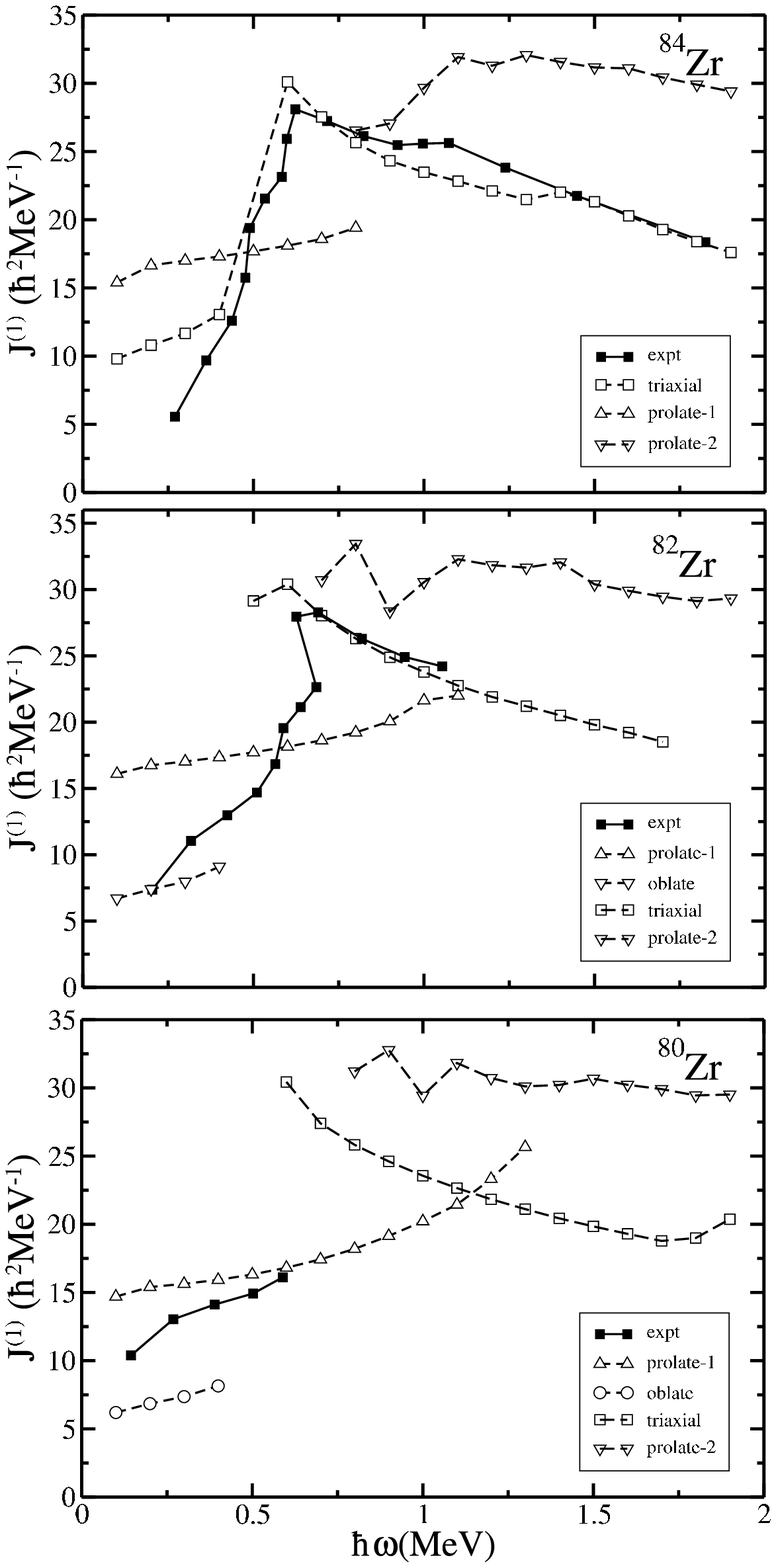}
\caption{Kinematic moments of inertia for $^{84}$Zr, $^{82}$Zr and
$^{80}$Zr as the function of the rotational frequency.}
\label{fig:4}
\end{figure}

\begin{figure}
\centering
\includegraphics[width=0.35\textwidth]{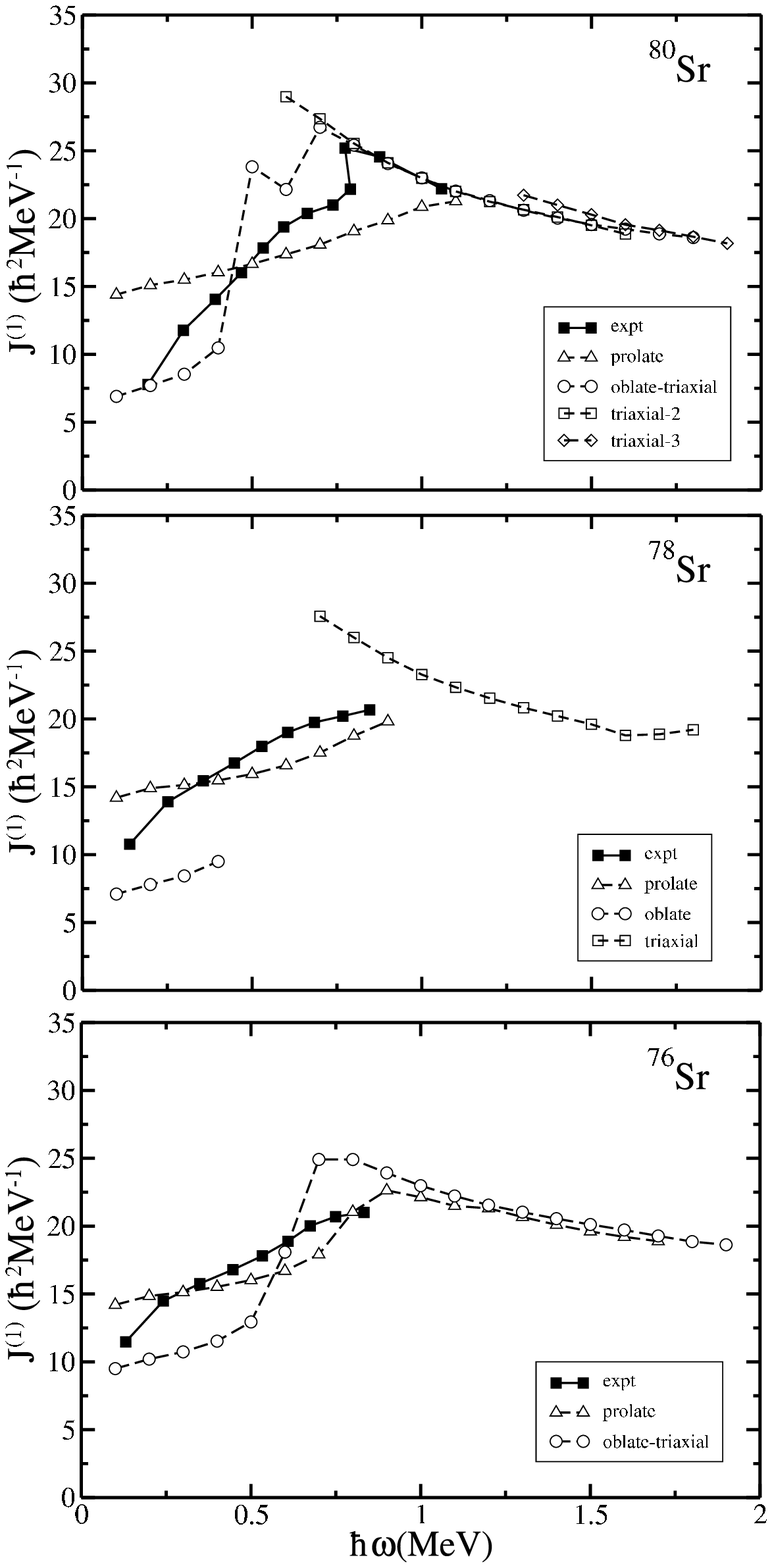}
\caption{Kinematic moments of inertia for $^{80}$Sr, $^{78}$Sr and
$^{76}$Sr as the function of the rotational frequency.
}
\label{fig:5}
\end{figure}

\begin{figure}
\centering
\includegraphics[width=0.35\textwidth]{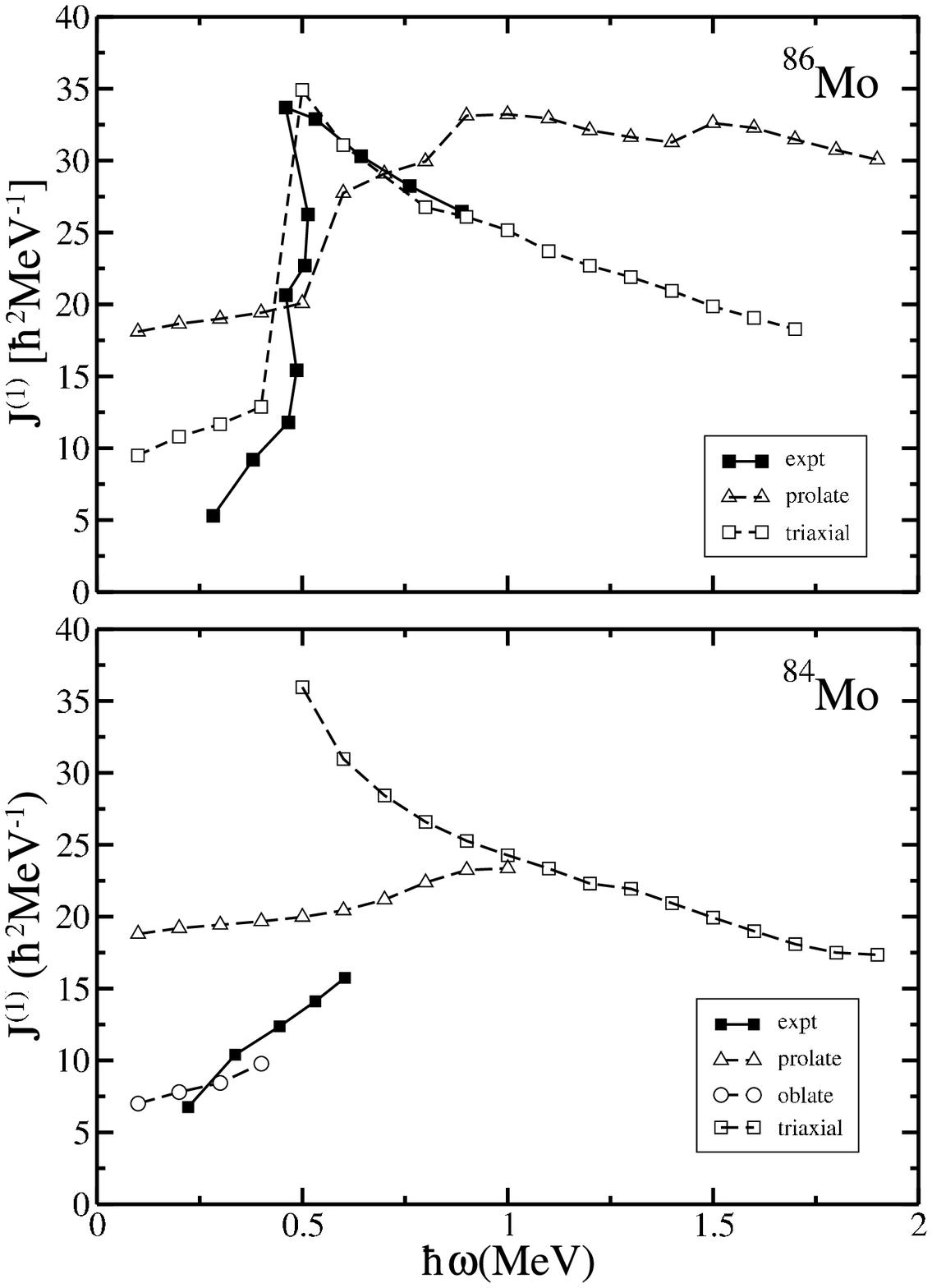}
\caption{Kinematic moments of inertia for $^{84}$Mo and $^{86}$Mo as
the function of the rotational frequency.
}
\label{fig:6}
\end{figure}

For the nucleus $^{84}$Zr, triaxial and prolate (prolate-1 and
prolate-2) deformations coexist in our calculations. As can be seen
from Fig.~\ref{fig:4}, the kinematic moments of inertia of the
triaxial rotational band agree rather well with the experimental
results. It shows a steep upbending at the rotational frequency
around 0.5 MeV/$\hbar$ due to the alignments of a pair of
quasiprotons and quasineutrons. After the alignments, a down slope
appears indicating the decreasing deformation which is also in
accordance with the triaxial deformation evolving. There are some
discrepancies between the experimental results and our calculations
at low spins for the nucleus $^{82}$Zr because the deformation is
considerably soft as discussed above. At higher spins after the band
crossing, a stable triaxial deformation appears and the resulting
kinematic moments of inertia of triaxiality agree well with the
experiment. Experimental data for $^{80}$Zr are limited and it is
impossible to show any trend at high rotational frequency. However,
the observed upbending behavior agrees well with the calculated
prolate band with frequency energy lower than 0.6 MeV. Triaxially
deformed band is predicted, which also exhibits a downbending
behavior and becomes yrast at higher rotational frequency.

Plots of the kinematic moments of inertia versus rotational
frequency for strontium and molybdenum isotopes shown in
Fig.~\ref{fig:5} and Fig.~\ref{fig:6} also show good consistence
between the experimental results and calculations. At high spins,
the triaxially deformed shapes appear and become stable after the
alignments. For the nucleus $^{80}$Sr, it is even complicated at
high spins. Varieties of shapes such as prolate, oblate-triaxial,
triaxial2 and triaxial3 shown in the Fig.~\ref{fig:5} coexist. They
have the nearly same moments of inertia and also the similar total
Routhians, indicating that several different configurations are
available at high spins. 
For the nucleus $^{84}$Mo, only a few experimental data are
available while it behaves like an oblate band at low spins. At the
rotational frequency of $\hbar\omega\approx0.5$ MeV, a shape
transition from oblate to triaxial deformation is predicted and it
will keep stable up to 24$\hbar$.


\section{summary}

In conclusion, the nuclei with mass number around 80, i.e.,
$^{76-80}$Sr, $^{80-84}$Zr and $^{84,86}$Mo have been calculated by
the TRS model. Due to the large gaps at the prolate deformation
$\beta_2\approx0.4$ with the particle number 38 or 40 and gaps at
oblate deformation $\beta_2\approx-0.3$ with the particle number 34
or 36, shape coexistences are prominent in this region. Spherical,
prolate, oblate and triaxial deformations coexist. Mostly, prolate
and oblate deformations coexist at low spins while the energy
difference between the two shapes deviates largely. For the nuclei
$^{76,78}$Sr and $^{78,80}$Zr, the Routhians of the prolate
deformations are lower than the oblate ones by more than 1 MeV,
while for the nuclei $^{80}$Sr and $^{82}$Zr, the energy differences
between the two deformations are less than 220 keV.

At high spins, due to the alignments of a pair of $g_{9/2}$ protons
and a pair of $g_{9/2}$ neutrons, triaxial deformations develop.
Both $^{82}$Zr and $^{84}$Zr show stable triaxial deformations after
the alignments. The triaxially deformed shape for $^{80}$Zr is
predicated to develop at high spins. The neighboring nuclei
$^{76-80}$Sr and $^{84,86}$Mo also show triaxiality at high spins.
Thus the $g_{9/2}$ orbits play a particularly role for the shape
transition of nuclei in this region.

\acknowledgments This work is supported by the Chinese Major State
Basic Research Development Program No. G2000077400, the National
Natural Science Foundation of China (Grant Nos. 10175002, 11065001,
and 10475002), the Doctoral Foundation of Chinese Ministry of
Education (20030001088), and by Suranaree University of Technology
under contract No. 15/2553.

\end{document}